\documentclass[a4paper,twocolumn,11pt,accepted=2021-06-15]{quantumarticle}
\pdfoutput=1
\usepackage[utf8]{inputenc}
\usepackage[english]{babel}
\usepackage[T1]{fontenc}
\usepackage{amsmath}
\usepackage{hyperref}
\usepackage{breqn}
\usepackage{tikz}
\usepackage{graphicx}
\usepackage{dcolumn}
\usepackage{bm}
\usepackage{hyperref}
\usepackage{float}
\usepackage{ragged2e}
\usepackage{hyperref}
\usepackage{float}

\usepackage[utf8]{inputenc}
\usepackage{amsmath}
\usepackage{mathrsfs}
\usepackage{float}
\usepackage{mathtools}

\usepackage{amsmath}

\begin{document}

\title{Self-referenced hologram of a single photon beam}

\author{Wiktor Szadowiak$^{\dag}$}
\affiliation{Institute of Experimental Physics, Faculty of Physics, University of Warsaw, ul. Pasteura 5,02-093 Warszawa, Poland}
\author{Sanjukta Kundu$^{\dag}$}
\affiliation{Institute of Experimental Physics, Faculty of Physics, University of Warsaw, ul. Pasteura 5,02-093 Warszawa, Poland}

\author{Jerzy Szuniewicz}
\affiliation{Institute of Experimental Physics, Faculty of Physics, University of Warsaw, ul. Pasteura 5,02-093 Warszawa, Poland}
\author{Radek Lapkiewicz*}
\orcid{0000-0002-2445-2701}
\affiliation{Institute of Experimental Physics, Faculty of Physics, University of Warsaw, ul. Pasteura 5,02-093 Warszawa, Poland}
\email{radek.lapkiewicz@fuw.edu.pl}
\thanks{\\$\dag$These authors contributed equally to this work.}
\maketitle

\begin{abstract}
  Quantitative characterization of the spatial structure of single photons is essential for free-space quantum communication and quantum imaging. We introduce an interferometric technique that enables the complete characterization of a two-dimensional probability amplitude of a single photon. Importantly, in contrast to methods that use a reference photon for the phase measurement, our technique relies on a single photon interfering with itself. Our setup comprises of a heralded single-photon source with an unknown spatial phase and a modified Mach-Zehnder interferometer with a spatial filter in one of its arms. The spatial filter removes the unknown spatial phase and the filtered beam interferes with the unaltered beam passing through the other arm of the interferometer. We experimentally confirm the feasibility of our technique by reconstructing the spatial phase of heralded single photons using the lowest order interference fringes. This technique can be applied to the characterization of arbitrary pure spatial states of single photons. 
\end{abstract}

\section{\label{sec:level1}Introduction}

The two-dimensional spatial phase profile of a single photon~\cite{spatial1,weak1,spatial2} is an important resource for free-space quantum optical communication~\cite{quancom4,quancom1,quancom2,quancom3,quancom5,quancomm,quancombom,optcom_new1}, quantum computing~\cite{quancomp0,quancomp}, quantum imaging~\cite{quanimg1,quanimg2,quanimg3,quanimg4,quanimg5,quanimg6,quanimg7,quanimg8} and quantum metrology~\cite{quanmetro2,quanmetro3,quanmetro4}. Unfortunately, the complete characterization of the two-dimensional spatial structure of a single photon is a challenging task due to its entirely indeterminate global phase~\cite{hologram,wignar}. Earlier, characterizing the spatial phase profile of single photons~\cite{photon_wavefn0,photon_wavefn1,photon_wavefn2} had been tackled using either tomographic techniques~\cite{tomo,tomo2,tomo3,tomo4} or weak measurement approaches~\cite{weak1,weak2,transverseSpatialStructure}. Nonetheless, these methods are not as straightforward and precise as the ones relying on interferometry~\cite{sagnac}.

In a recent work~\cite{hologram}, difficulties of retrieving the spatial phase profile of a single photon beam have been overcome ingeniously by using Hong-Ou-Mandel interference~\cite{hongoumandel}. The unknown photon with an arbitrary spatial phase profile was overlapped with a reference photon of a constant phase profile. In that scheme, the reference photon has to be identical in all of its degrees of freedom, except the spatial one, to the photon of unknown spatial phase. Therefore, the experimental preparation of the reference photon becomes challenging. Our method, which is analogous to a classical holography technique~\cite{diff_microscopy, microprin, optholo}, offers a way of characterizing the spatial phase of the single-photon beam in spite of its indeterminate global phase. Similarly to \cite{hologram}, by holography we mean a measurement of the phase profile of a light beam using interference \cite{OffaxisHolo}. Preliminary results of this work have been presented at conferences~\cite{FiO,Rochester}. Contrary to the previous approach~\cite{hologram} and its recent generalization to mixed states~\cite{hologramGen}, which use two-photon interference~\cite{twophotonInterf1}, we use single-photon interference~\cite{twophotonInterf1,singlePhotonInterference1}.
\begin{figure}[ht!]
\centering
    \captionsetup{width=.99\linewidth}
    \centering
    \includegraphics[width=\linewidth]{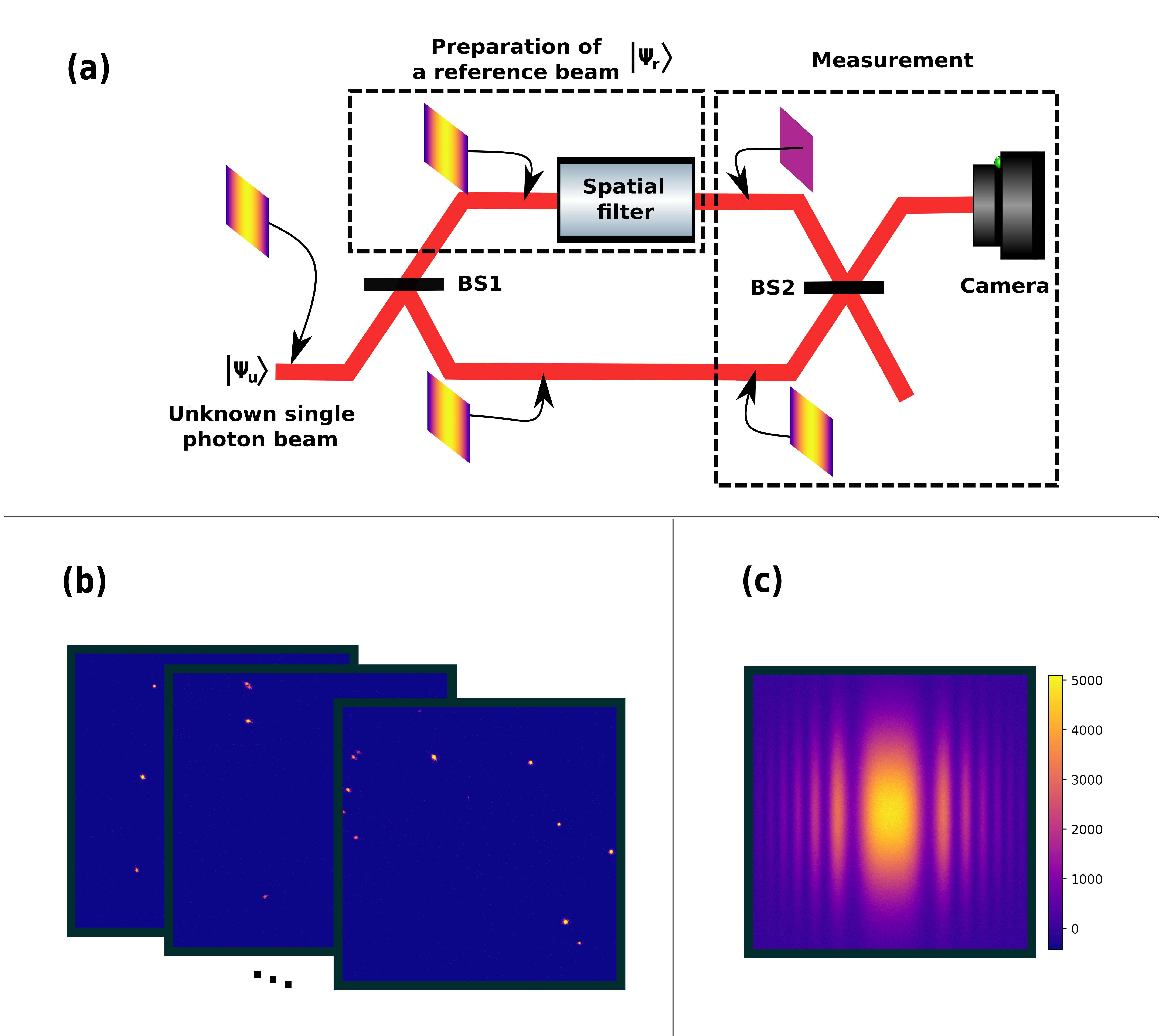}
   \caption[0.98\linewidth]{\textbf{(a)} Schematic representation of the experiment. An incoming beam of single photons with an unknown spatial phase profile is split at a beam splitter BS1 and the resulting beams go through the two arms of a Mach-Zehnder interferometer. In the upper arm, the spatial structure of the photons is erased when the beam passes through a spatial filter which produces the reference beam $\psi_r(x,y)$. The unknown and reference beams are combined at beam splitter BS2 and photons are registered with the help of the single photon sensitive camera. The spatial phase profile of the unknown beam is recovered from the recorded interferogram. \textbf{(b)} Stack of captured frames with single photon detection events visible as bright spots. \textbf{(c)} The interferogram (simulated) is obtained by creating a 2D histogram of the positions of photodetections.}
\label{fig:schem}
\end{figure}
\begin{figure*}[ht!]
  \centering
  \captionsetup{width=\linewidth}
  \includegraphics[width=\linewidth]{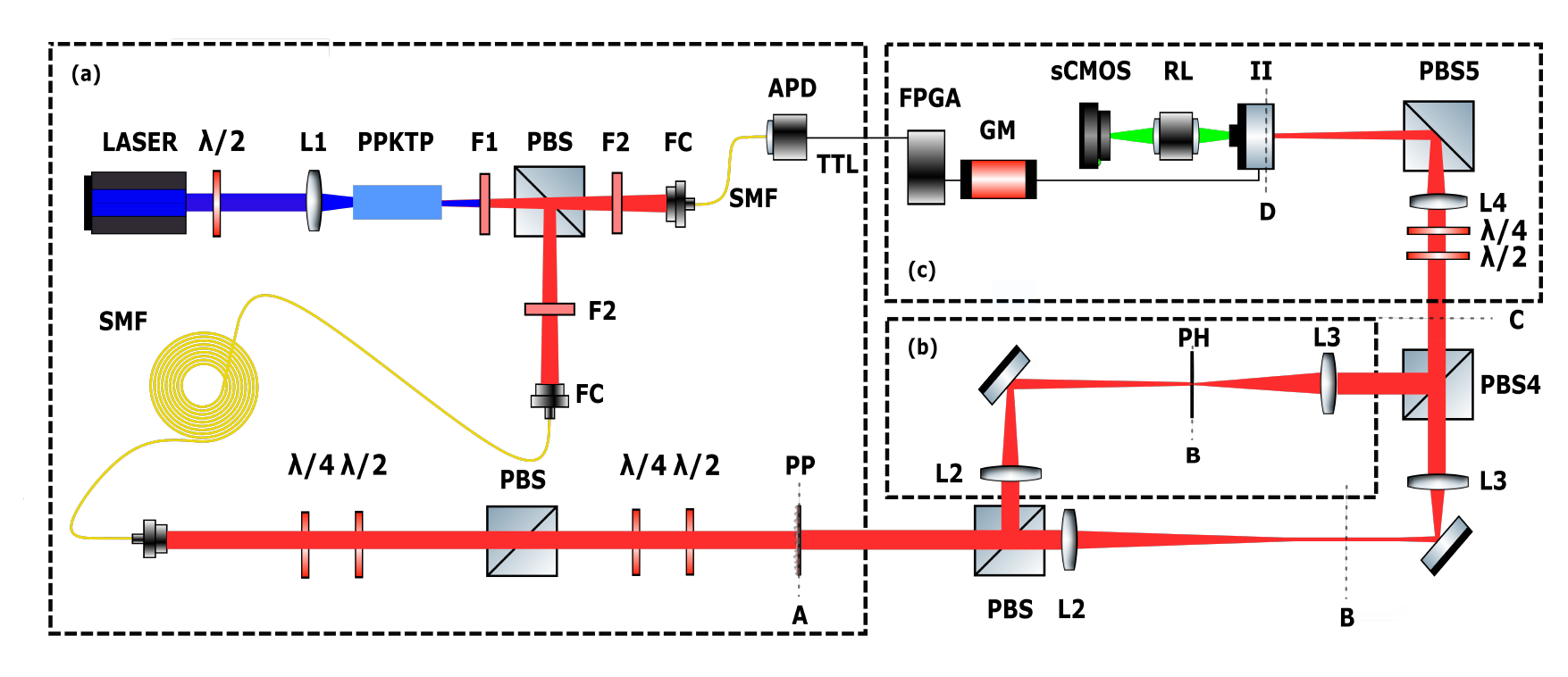}
\caption{ Experimental set-up for a self-referenced measurement of a spatial structure of a single photon beam with an unknown phase profile. \textbf{(a)}  Preparation of an unknown single photon beam $\psi_u(x,y)$. Orthogonally polarized collinear photons signal and idler generated in SPDC type-II process in a Nonlinear PPKTP crystal, L-lens with focal length of 300mm; after passing through a F1-long pass filter get separated by the PBS - polarization beam splitter, F - filter transmitting 810 nm $\pm$ 5 nm, FC - optical fibre coupler; Photons are transmitted through the SMF - single-mode fibre, $\frac{\lambda}{2}$-half wave plate, L1 - lens with focal length of 150 mm, $\frac{\lambda}{4}$-quarter wave plate, PP (plane A) - phase imprinting plane, APD - Avalanche photo-diode; \textbf{(b)} Preparation of the reference beam $\psi_r(x,y)$. M - mirrors; L2, L3 - lenses with focal lengths of 250 mm, 125mm respectively; PH - pinhole of size 200 $\mu$m, plane B - intermediate plane of 4f system; plane C -image plane of 4f system \textbf{(c)} Measurement.  IsCMOS - Intensified sCMOS camera, it consists of a II - image intensifier, RL - Relay lens and a sCMOS camera (Andor Zyla 4.2). Plane D - Image plane on the camera, FPGA - field-programmable gate array, GM - gating module.}
\label{fig:Expt_setup}
\end{figure*}

\section{Concept}
We present an interferometric system, illustrated in Fig.\,1a, to reconstruct the unknown spatial phase $\phi(x,y)$ of a single photon beam characterized by the probability amplitude, $\psi_u(x,y)=|\psi_u(x,y)| e^{i\phi(x,y)}$. In one of the arms of the Mach-Zehnder interferometer, we place a spatial filter so that the photon after filtering carries only a flat phase profile (not dependent on position). In this way we create the reference beam characterized by a probability amplitude $\psi_r(x,y) = |\psi_r(x,y)|e^{i\phi_0}$, where $\phi_0$ is a constant phase. The unknown and the reference beams are overlapped at a beamsplitter BS2. We subsequently detect the positions of the single photons with a spatially resolving detector and after accumulating many photodetections (Fig.\,1b), we obtain the interferogram (Fig.\,1c):\\
\begin{align}\label{eq1}
|\psi(x,y)|^2 & = |\psi_u(x,y) + \psi_r(x,y)|^2\nonumber\\
 & =  |\psi_u(x,y)|^2 + |\psi_r(x,y)|^2 \\ &+2|\psi_u(x,y)||\psi_r(x,y)|\cos({\phi(x,y)-\phi_0}).\nonumber
\end{align}

Note that, the last term in the above equation contains the information of the spatial phase profile $\phi(x,y)$ of the unknown photon.
We reconstruct the unknown phase of the single-photon beam by using Fourier off-axis holography~\cite{OffaxisHolo} (see the supplementary material).

\section{Experimental design}
The experimental setup is depicted in the Fig.\,2. Heralded single photons are prepared via collinear, type-II spontaneous parametric down conversion (SPDC) process~\cite{spdc2} in a PPKTP nonlinear crystal pumped with a 405 nm laser beam of approximately 120 mW from a continuous wave diode laser. Both signal and idler photons generated as a pair are of the same central wavelength (810 nm) but orthogonal polarization. The two photons are separated at a polarizing beam splitter (PBS), spectrally filtered by an interference filter of 3 nm FWHM, and spatially filtered by coupling into a single mode fiber (SMF). An avalanche photo-diode (APD) registers one of the photons from the pair produced in SPDC and generates a short TTL pulse. This pulse is used to trigger our custom-built intensified sCMOS (IsCMOS) camera~\cite{IsCMOS} using a gating module (Photek GM300-3N) with a mean trigger rate of 200 kHz to herald the other photon from the pair which we use as our heralded signal. The signal photon is then transmitted through a 20 m long spool of a single mode fiber to synchronize the gating time of the image intensifier and the arrival of the signal photon by compensating the processing time of the APD and IsCMOS.

In order to prepare the single photon beam with an unknown spatial phase profile, we place a phase mask in plane A (Fig.\,2a.). We perform several experiments with different phase masks applied to the single photon beam. We imprint a 2D quadratic phase and a spiral phase by placing a spherical lens with a focal length of 125 mm and a spiral phase plate respectively in the plane A. Similarly we imprint a 1D quadratic phase profile to the single photon beam by placing a cylindrical lens of focal length 75 mm in a distance of 17 mm before plane A. In this case, due to the extra propagation, the phase profile in the plane A is equivalent to the phase created by a cylindrical lens with a focal length of 58 mm.

The beam of single photons with a spatial phase enters the Mach-Zehnder interferometer. Both the arms of the interferometer consist of 4f imaging systems using lenses L2, L3 with a $1.6$ magnification factor, where the object and the image planes of the 4f-systems of both arms are the planes A and C respectively. In one arm, lenses L2 and L3 with a pinhole placed in plane B, the Fourier plane of the 4f system~\cite{4f}, are used as a spatial filter removing higher spatial frequencies. After passing through the pinhole, the beam can be approximated by a Gaussian beam and can act as a reference beam. 

For some phases (e.g., the spiral phase), the beam to be characterized contains only higher spatial frequencies and because of that, the pinhole would block all the light in the filter arm. In such a situation, the pinhole has to be shifted transversely to maximize the transmission through the arm with the spatial filter. As a result of the shift, the reference beam acquires an extra linear phase which does not affect the reconstructed phase.

\begin{figure*}[ht!]
\centering

  \includegraphics[width=\linewidth]{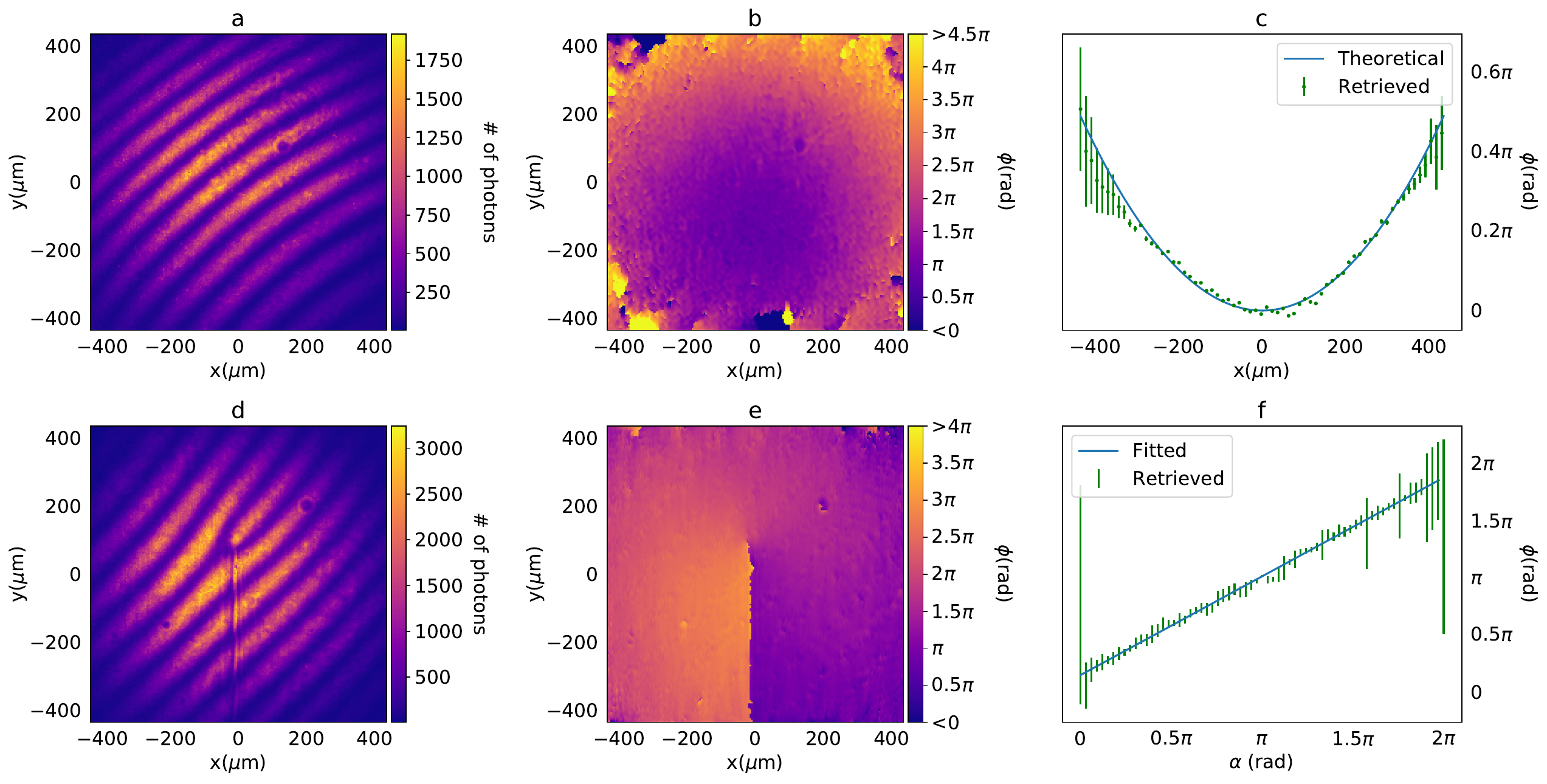}
   \usebox0
\caption{\textbf{(a)} the recorded interferogram for 2D quadratic  phase; \textbf{(b)} the reconstructed phase for 2D quadratic  phase ; \textbf{(c)} plots of the theoretical quadratic phase (blue solid line) and the reconstructed phase (dots) averaged over middle 50 columns of the plot in \textbf{b} with standard deviation in case of spherical phase mask applied. In the middle part, standard deviation is smaller than $\frac{2\pi}{150}$; \textbf{(d)} the recorded interferogram for spiral phases; \textbf{(e)} the  reconstructed phase for spiral phase; \textbf{(f)} plots of the reconstructed phase in case of spiral phase mask with standard deviations versus the azimuthal angle $\alpha$ and a fitted linear function.}
\label{fig:sph_oam}
\end{figure*}

The polarizations of the reference and signal beams are orthogonal. Therefore, in order to observe interference of these beams we overlap them spatially at PBS4, rotate their polarization using waveplates, and project their polarization onto a linear polarization using PBS5 (Fig.\,2b,c). Simultaneously, plane C is imaged by L4 lens onto the plane D -- the photo-cathode of the image intensifier. Each of the photons, that illuminate the image intensifier, has a chance to eject a photoelectron which produces a macroscopic charge avalanche, which in turn generates a bright flash at the phosphor screen located at the back of the image intensifier. These bright flashes are finally imaged onto the sCMOS sensor by a relay lens composed of two photographic objectives (Fig.\,2c.). 
We use our camera in the photon counting mode in which, we fit a Gaussian function to each flash detected by the sCMOS camera and store its central position~\cite{softwarePhotonfinder}. By this method, we can eliminate the readout noise of the sCMOS sensor and achieve sub-pixel spatial resolution.

In our experiment, the sCMOS sensor is set to have an acquisition rate of 9Hz with a region of interest of 1000 px$\times$1000 px. The image intensifier is opened for approximately 20 ns after each trigger. In the case of two trigger signals appearing closer than 3 $\mu$s apart, the gating module starts operating randomly independent of the appearing signal rate. Therefore, in order to meet the technical requirements of the gating module, we introduce an FPGA that discriminates the second signal whenever there are two triggers arriving very close to each other. This setup lets us maximize the photon count rate and ensure that the photons detected by the camera are indeed heralded. In our measurement, the rate of dark counts is around 25 Hz. Another source of noise can be accidental coincidence counts, i.e. detection of photons that are not heralded but reach the image intensifier during the gate opening time. The accidental coincidence rate is up to 3 Hz. It is, in fact, possible to increase the SNR by using an image intensifier that allows for shorter gate lengths since this results in minimizing the dark counts as well as accidental coincident counts without reducing the signal level.
\begin{figure*}[ht!]
\captionsetup{width=.95\linewidth}
  \includegraphics[width=\linewidth]{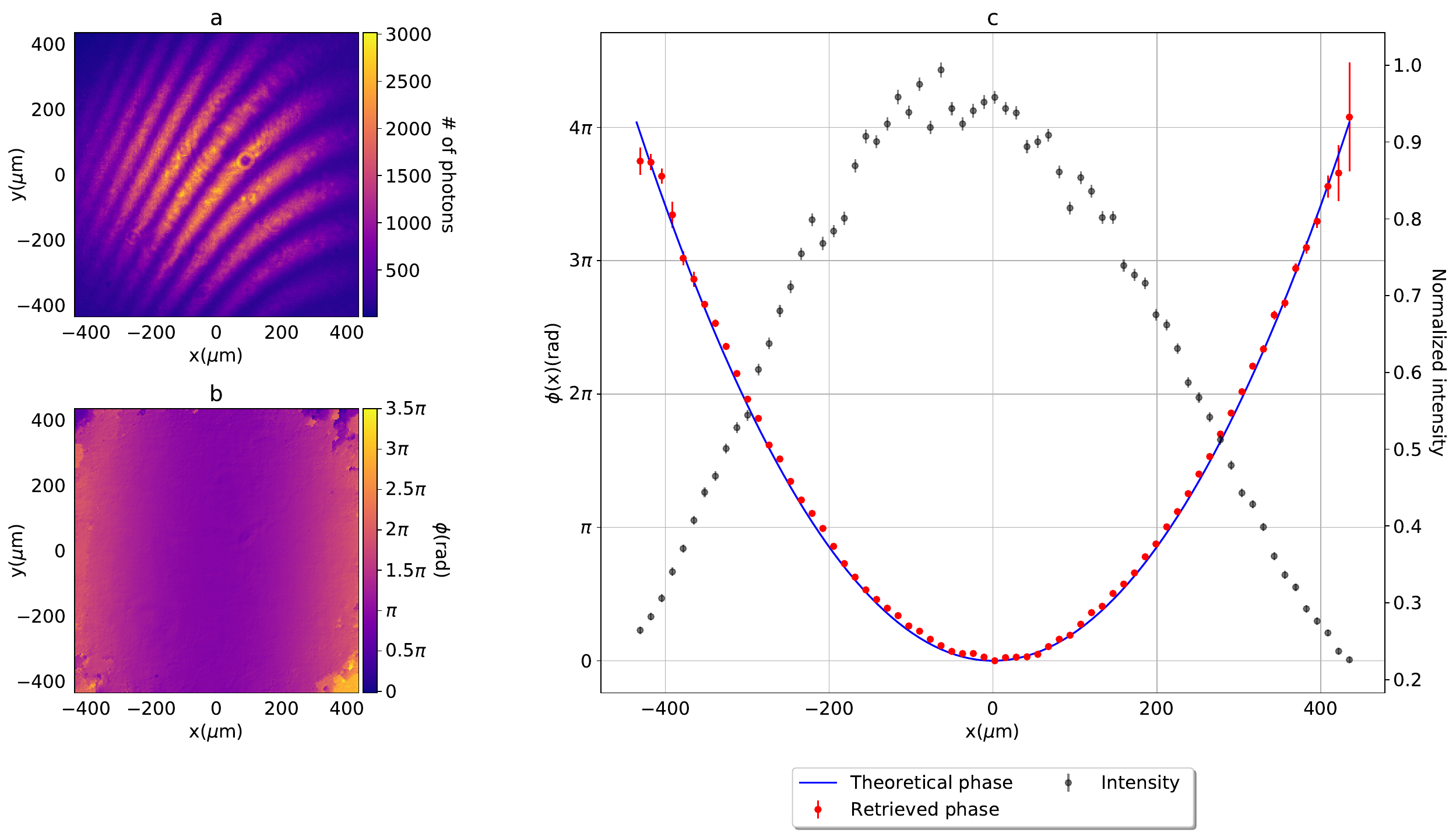}
\caption{\textbf{(a)} Measured interferogram with 1D quadratic phase, \textbf{(b)} Retrieved phase, \textbf{(c)} Retrieved phase and amplitude with uncertainty and error-bars. The error-bars are scaled to 3 times their actual value for the retrieved phase and 5 times for intensity. There is a theoretical plot for comparison of retrieved phase based on $\phi(x)=\frac{\pi x^2}{\lambda f}$; $\lambda$ is the wavelength of the photon and $f$ is the focal length of the cylindrical lens used as a 1D quadratic phase object. In all the plots, x and y denote the positions at the plane A (Fig.\,2).}
\label{fig:cyl_ph_amp}
\end{figure*}
\section{Results}

In our experiment, in order to  prepare beams with unknown spatial phase profiles, we insert the following phase masks into a heralded single photon beam: a spherical lens, a spiral phase plate, and a cylindrical lens. The quantitative analysis for reconstructing the spatial phase profiles of the beams is then realized using separate measurements for each of the aforementioned cases. This process is carried out by examining the interferograms that result from registering a total of $1\times10^7-2.5\times 10^7$ heralded single photons on an intensified sCMOS camera~\cite{IsCMOS}, with about 200 to 600 photons being detected in 1 s.

The recorded interferograms for 2D quadratic and spiral phases are shown in the first column of Fig.\,3 (Fig.\,3a and Fig.\,3d) while their corresponding 2D-reconstructed phases are presented in the second column of Fig.\,3 (Fig.\,3b and Fig.\,3e). The details of the phase reconstruction algorithm have been described in the Supplementary information. In order to estimate the statistical uncertainties of the reconstructed phase and intensity, we use a simulation: We repeat the phase retrieval process 1000 times, each time drawing the photon number at every pixel from a Poisson distribution with the mean given by the number of photons detected at this pixel. From all of the 1000 reconstructed phases, we are finally able to estimate the mean and the uncertainty of the phase at each pixel. The single pixel uncertainties of the reconstructed 2D phases are below $\frac{2\pi}{30}$ and $\frac{2\pi}{50}$ for the 2D quadratic phase and spiral phase in the region with a radius of 50 pixels with maximum visibility and photon count rate. In Fig.\,3c, we plot the values of the reconstructed 2D quadratic phase averaged over the middle 50 columns of Fig.\,3b with the statistical uncertainties obtained from the numerical simulation. The theoretical plot is also shown (Fig.\,3c - blue line). In the middle part, the standard deviation is smaller than $\frac{2\pi}{150}$. Here the plot shows an excellent agreement with the theory although the accuracy of our technique is limited by the systematic errors rather than the statistical uncertainties. As expected for the spiral phase plate, Fig.\,3f shows that the reconstructed phase depends linearly on the azimuthal angle.

Lastly, Fig.\,4a and Fig.\,4b represent the recorded interferogram and the corresponding reconstructed phase when 1D quadratic phase mask is applied. The single pixel uncertainty of the reconstructed 2D phase Fig.\,4b is below $\frac{2\pi}{50}$ in the region with a radius of 50 pixels. Fig.\,4c shows the plots of the reconstructed phase and intensity profile of the single photon beam. Each point here represents an average over 50 pixels from the middle of Fig.\,4b. In addition to that, we plot the theoretical phase (Fig.\,4c -- blue line), which in the case of the 1D quadratic phase can be represented by $\phi(x)=\frac{\pi x^2}{\lambda f}$, where $\lambda$ is the wavelength of the photon and $f$ is the focal length of the cylindrical lens which is used as a 1D quadratic phase object. The plot in Fig.\,4c shows good agreement between the theoretical and our reconstructed phase. The data analysis requires an extra linear phase to be introduced between the reference and unknown beam in the interferometer (See Supplementary information). The interference visibilities in the case of 2D quadratic, spiral and 1D quadratic phases are approximately 69\%, 78\%, and 77\% respectively.

\section{Discussion}

In summary, we introduced a self-referenced interferometeric technique that can be used to reconstruct both the spatial phase and the amplitude profile of an unknown single photon beam. Our method does not require a reference photon which makes our approach robust and straightforward to implement. 
Moreover, in contrast to methods relying on the measurements of a two-photon probability distribution (e.g. \cite{hologram}), our technique does not require coincidence detection. As a consequence, self-referenced holography is significantly simpler as far as the technical aspect is concerned and is more tolerant to losses. In contrast to ~\cite{hologram} the current implementation of our method is susceptible to interferometric phase fluctuations. This difficulty could be eliminated by using common path designs~\cite{diff_microscopy}. In the case of applications in which the loss introduced by the spatial filter cannot be tolerated, shearing interferometry~\cite{shearing1} could be used instead of spatially filtered reference.

We demonstrated that our method is not limited to only one-dimensional phase reconstruction and can characterize any single photon beam with an unknown pure spatial state including orbital angular momentum modes~\cite{oam1,oam2}. Because of its simplicity, this measurement scheme has the potential for multiple applications in the fields of quantum communication and imaging. Similar techniques can be used to characterize the spectral degrees of freedom of photons~\cite{SPIDER}. In addition, analogous techniques might also find applications in matter wave interferometry~\cite{matterWave,matterWave2}. It would be interesting to study whether our method might be generalized to mixed~\cite{hologramGen} or entangled states.

\section{Funding}
National Science Centre (Poland) Grant No. 2015/17/D/ST2/03471; the Polish Ministry of Science and Higher Education; Foundation for Polish Science under the FIRST TEAM project ‘Spatiotemporal photon correlation measurements for quantum metrology and super-resolution microscopy’ co-financed by the European Union under the European Regional Development Fund Grant No. POIR.04.04.00-00-3004/17-00; National Laboratory for Photonics and Quantum Technologies (NLPQT) Grant No. POIR.04.02.00.00-B003/18. 

\section{Acknowledgments}
We thank W. Wasilewski, K. Banaszek, P. Fita, R. Chrapkiewicz, M. Jachura and G. Firlik for the support with the home-built intensified camera, discussions and equipment loans and M. Lahiri, R. Fickler, B. Ghosh and T. Chanda for discussions about the manuscript.

\section{Data and code availability}
The data and the codes used to process the data are available from the corresponding author upon reasonable request.

\medskip

\noindent\textbf{Disclosures.} The authors declare no conflicts of interest.
\medskip

\noindent See Supplementary for supporting content.

\onecolumn
\pagebreak
\renewcommand{\thesection}{S\arabic{section}}
\renewcommand{\thesection}{S\arabic{subsection}}

\numberwithin{equation}{section} 
\renewcommand{\theequation}{Eq. \thesection.\arabic{equation}} 

\section{Supplementary material: Fourier off-axis holography}\label{FourierOFF}



In our experiment, linear phase is introduced in the interferometer by tilting the beams with respect to each other. As a result they overlap only at the 4f image plane and the final image plane (planes C and D in Fig. 2). The goal of introducing a tilt between the interfering beams is to add a linear phase to the interferogram, which is preferred for the data analysis. A reference measurement of the linear phase which needs to be subtracted at the end of the phase reconstruction, simplifies the analysis of the measured interferograms.

We begin with the first order interferogram recorded by collecting the positions of the accumulated photons
\begin{align}\label{eq1}
|\psi(x,y)|^2 & = |\psi_u(x,y) + \psi_r(x,y)|^2\nonumber\\
 & =  |\psi_u(x,y)|^2 + |\psi_r(x,y)|^2 \\ &+2|\psi_u(x,y)||\psi_r(x,y)|\cos({\phi'(x,y)-\phi_0}),\nonumber
\end{align}
where $\phi_0$ is the interferometric phase offset and the spatial phase between the interfering beams $\phi'(x,y)$ can be represented as
\begin{equation}
\phi'(x,y) = (a_{1}x+a_{2}y)+\phi(x,y),
\label{eq:refname4}
\end{equation}
where $(a_{1}x+a_{2}y)$ denotes the linear phase introduced in the interferometer, $\phi(x,y)$ represents an arbitrary phase that we want to reconstruct, and the constant phase is neglected.
The first step in the analysis is the 2D Fourier transform of the collected data. In accordance with the properties of the Fourier transform, the addition of a linear phase results in splitting the phase-dependent part of the signal with respect to the center of the Fourier plane.

Fourier transform of Eq. (\ref{eq1}) gives
\begin{equation}\label{eq:refname7}
\begin{split}
 \mathscr{F}[|\psi(x,y)|^2] &= \mathscr{F}[|\psi_u(x,y)|^2 + |\psi_r(x,y)|^2]\\
  & +\mathscr{F}[p(x,y)](k_{x}+a_{1},k_{y}+a_{2})\\
  & +\mathscr{F}[p^{*}(x,y)](k_{x}-a_{1},k_{y}-a_{2}),   
\end{split}
\end{equation}
where $p(x,y)=|\psi_u(x,y)|\times|\psi_r(x,y)|\times \exp{(-i(\phi(x,y)-\phi_0))}$ and the first term on the right hand side contains only phase-independent parts.
Eq. (\ref{eq:refname7}) reflects the fact that both of the phase-dependent terms convey the same information about phase to be retrieved.

The value of the linear phase has to be high enough to enable the separation of the three parts of Eq. (\ref{eq:refname7}) in the Fourier domain. By filtering in the Fourier domain we are able to retain only one of the three terms from Eq. (\ref{eq:refname7}). Note that, we do not lose any information about the phase. As a result of this analysis, we end up with $\mathscr{F}[p(x,y)](k_{x}+a_{1},k_{y}+a_{2})$ whose inverse Fourier transform leads to $|\psi_u(x,y)|\times|\psi_r(x,y)|\times\exp(i(\phi'(x,y)-\phi_0))$. It is then possible to retrieve phase $\phi'(x,y)$ by extracting the argument of the complex number of each pixel. To calculate the final result from the wrapped phase we use the 2D phase-unwrapping algorithm \cite{unwrap}. It should be kept in mind that the extra linear phase has to be removed. In order to do so, we repeat the whole procedure for the reference measurement with no phase mask and thus retrieve the linear phase, $(a_{1}x+a_{2}y)$. By subtracting the reference linear phase and disregarding the constant phase offset $\phi_0$, we finally obtain the desired imprinted phase $\phi(x,y)$.

Our reconstruction gives the best results for high visibility interference fringes, which require matching the intensity profiles of reference and unknown beams. This requirement is valid for most interferometric techniques, and in our experiment it is satisfied by using tunable beamsplitters and matching the profiles of the reference beam and the unknown beam. This beam profile matching is performed by checking the intensity profiles of both beams separately with a single-photon sensitive camera. One could also ensure that the unknown and reference beams’ intensity profiles are matched by maximizing the visibility of interference fringes.
Separate measurements of both beams' intensity profiles allow for tuning the splitting ratios of both beamsplitters (BS1 and BS2 in Fig.\,1a.). This is necessary for overcoming the variable loss introduced by the spatial filter, depending on the object’s phase. 



\medskip
\begin{thebibliography}{99} 

\bibliographystyle{unsrt}
\bibitem{spatial1} G. M. Terriza, J. P. Torres and L. Torner, Twisted photons, Nature Physics \textbf{3}, 305–310 (2007).
\href{https://doi.org/10.1038/nphys607}{DOI: 10.1038/nphys607.}

\bibitem{weak1} J. S. Lundeen, B. Sutherland, A. Patel, C. Stewart \& C. Bamber, Direct measurement of the quantum wavefunction, Nature \textbf{474}, 188–191 (2011).
\href{https://doi.org/10.1038/nature10120}{DOI: 10.1038/nature10120.}


\bibitem{spatial2} R. Fickler, R. Lapkiewicz, W. N. Plick, M. Krenn, C. Schaeff, S. Ramelow, A. Zeilinger, Quantum entanglement of high angular momenta, Science \textbf{338}, 640–643 (2012).
\href{https://doi.org/10.1126/science.1227193}{DOI: 10.1126/science.1227193.}

\bibitem{quancom4} M. Mirhosseini, M. Malik, Z. Shi, and R. W. Boyd, Efficient separation of
the orbital angular momentum eigenstates of ligh, Nat. Commun. \textbf{4}, 2781 (2013).
\href{https://doi.org/10.1038/ncomms3781}{DOI: 10.1038/ncomms3781.}


\bibitem{quancom1} S. Walborn, D. Lemelle,  M. Almeida, P. Ribeiro, Quantum key distribution with higher-order alphabets using spatially encoded qudits, Phys. Rev. Lett. \textbf{96}, 090501 (2006).
\href{https://doi.org/10.1103/PhysRevLett.96.090501}{DOI: 10.1103/PhysRevLett.96.090501.}

\bibitem{quancom2} S. F. Pereira, Z. Y. Ou, and H. J. Kimble, Quantum communication with correlated nonclassical states, Phys. Rev. A \textbf{62}, 042311 (2000).
\href{https://doi.org/10.1103/PhysRevA.62.042311}{DOI: 10.1103/PhysRevA.62.042311.}

\bibitem{quancom3} R. Fickler, R. Lapkiewicz, M. Huber, M. P.J. Lavery, M. J. Padgett \& A. Zeilinger, Interface between path and orbital angular momentum entanglement for high-dimensional photonic quantum information, Nature Communications \textbf{5}, 4502 (2014).
\href{https://doi.org/10.1038/ncomms5502}{DOI: 10.1038/ncomms5502.}

\bibitem{quancom5} M. Aspelmeyer, T. Jennewein, M. Pfennigbauer, W. R. Leeb, A. Zeilinger, Long distance quantum communication with entangled photons using satellites, IEEE J. Sel. Top. Quantum Electron \textbf{9}, 1541 (2003).
\href{https://doi.org/10.1109/JSTQE.2003.820918}{DOI: 10.1109/JSTQE.2003.820918.}

\bibitem{quancomm} M. Krenn, J. Handsteiner, M. Fink, R. Fickler, and A. Zeilinger. Twisted photon entanglement through turbulent air across Vienna, Proc. Natl Acad. Sci. USA \textbf{112}, 14197–14201 (2015).
\href{https://doi.org/10.1073/pnas.1517574112}{DOI: 10.1073/pnas.1517574112.}

\bibitem{quancombom} O. Lib, G. Hasson, Y. Bromberg, Real-time shaping of entangled photons by classical control and feedback, Science Advances \textbf{6}, eabb6298 (2020). 
\href{https://doi.org/10.1126/sciadv.abb6298}{DOI: 10.1126/sciadv.abb6298.}

\bibitem{optcom_new1} H. Sasada and M. Okamoto, Transverse-mode beam splitter of a light beam and its application to quantum cryptography, Phys. Rev. A \textbf{68}, 012323 (2003).
\href{https://doi.org/10.1103/PhysRevA.68.012323}{DOI: 10.1103/PhysRevA.68.012323.}


\bibitem{quancomp0} S. Leedumrongwatthanakun, L. Innocenti, H. Defienne, T. Juffmann, A. Ferraro, M. Paternostro, S. Gigan, Programmable linear quantum networks with a multimode fibre, Nat. Photonics \textbf{14}, 139–142 (2020).
\href{https://doi.org/10.1038/s41566-019-0553-9}{DOI: 10.1038/s41566-019-0553-9.}


\bibitem{quancomp} A. F. Abouraddy, G. Di Giuseppe, T. M. Yarnall, M. C. Teich, and B. E. A. Saleh, Implementing one-photon three-qubit quantum gates using spatial light modulators, Phys. Rev. A \textbf{86}, 050303(R) (2012).
\href{https://doi.org/10.1103/PhysRevA.86.050303}{DOI: 10.1103/PhysRevA.86.050303.}

\bibitem{quanimg1} T. Ono, R. Okamoto \& S. Takeuchi, An entanglement-enhanced microscope, Nature Communications \textbf{4}, 2426 (2013).
\href{https://doi.org/10.1038/ncomms3426}{DOI: 10.1038/ncomms3426.}

\bibitem{quanimg2} P. A. Moreau, E. Toninelli, T. Gregory \& M. J. Padgett, Imaging with quantum states of light, Nature Reviews Physics \textbf{1}, 367–380 (2019).
\href{https://doi.org/10.1038/s42254-019-0056-0}{DOI: 10.1038/s42254-019-0056-0.}


\bibitem{quanimg3} Z. I. Borja, C. S. Gutiérrez, R. R. Alarcón, H. C. Ramírez, and A. B. U’Ren, Experimental demonstration of full-field quantum optical coherence tomography, Photon. Res. \textbf{8}, 51-56 (2020).
\href{https://doi.org/10.1364/PRJ.8.000051}{DOI: 10.1364/PRJ.8.000051.}

\bibitem{quanimg4} R. S. Aspden, D. S. Tasca, R. W. Boyd \& M. J. Padgett, EPR-based ghost imaging using a single-photon-sensitive camera, New J. Phys. \textbf{15}, 073032 (2013).
\href{https://doi.org/10.1088/1367-2630/15/7/073032}{DOI: 10.1088/1367-2630/15/7/073032.}

\bibitem{quanimg5} G. B. Lemos, V. Borish, G. D. Cole, S. Ramelow, R. Lapkiewicz \& A. Zeilinger, Quantum imaging with undetected photons, Nature \textbf{512}, 409–412 (2014).
\href{https://doi.org/10.1038/nature13586}{DOI: 10.1038/nature13586.}

\bibitem{quanimg6} A. F. Abouraddy, P. R. Stone, A. V. Sergienko, B. E. A. Saleh, and M. C. Teich, Entangled-Photon Imaging of a Pure Phase Object, Phys. Rev. Lett. \textbf{93}, 213903 (2004).
\href{https://doi.org/10.1103/PhysRevLett.93.213903}{DOI: 10.1103/PhysRevLett.93.213903.}

\bibitem{quanimg7} R. Fickler, M. Krenn, R. Łapkiewicz, S. Ramelow and A. Zeilinger, Real-time imaging of quantum entanglement, Sci. Rep. \textbf{3}, 1914 (2013).
\href{https://doi.org/10.1038/srep01914}{DOI: 10.1038/srep01914.}


\bibitem{quanimg8} P. A. Morris, R. S. Aspden, J. E. C. Bell, R. W. Boyd and M. J. Padgett, Imaging with a small number of photons, Nature Commun. \textbf{6}, 5913 (2015).
\href{https://doi.org/10.1038/ncomms6913}{DOI: 10.1038/ncomms6913.}


\bibitem{quanmetro2} V. Giovannetti, S. Lloyd, and L. Maccone, Quantum Metrology, Phys. Rev. Lett. \textbf{96}, 010401 (2006).
\href{https://doi.org/10.1103/PhysRevLett.96.010401}{DOI: 10.1103/PhysRevLett.96.010401.}

\bibitem{quanmetro3} M. Parniak, S. Borówka, K. Boroszko, W. Wasilewski, K. Banaszek, and R. D. Dobrzański, Beating the Rayleigh Limit Using Two-Photon Interference, Phys. Rev. Lett. \textbf{121}, 250503 (2018).
\href{https://doi.org/10.1103/PhysRevLett.121.250503}{DOI: 10.1103/PhysRevLett.121.250503.}

\bibitem{quanmetro4} M. Tsang, R. Nair, and X. Lu, Quantum Theory of Superresolution for Two Incoherent Optical Point Sources, Phys. Rev. X \textbf{6}, 031033 (2016).
\href{http://doi.org/10.1103/PhysRevX.6.031033}{DOI: 10.1103/PhysRevX.6.031033.}




\bibitem{hologram} R. Chrapkiewicz, M. Jachura, K. Banaszek and W. Wasilewski, Hologram of a single photon, Nat. Photonics \textbf{10}, 576–579 (2016).
\href{https://doi.org/10.1038/nphoton.2016.129}{DOI:
10.1038/nphoton.2016.129.}


\bibitem{wignar} A. I. Lvovsky, H. Hansen, T. Aichele, O. Benson, J. Mlynek, and S. Schiller, Quantum state reconstruction of the single-photon Fock state, Phys. Rev. Lett. \textbf{87}, 050402 (2001).
\href{https://doi.org/10.1103/PhysRevLett.87.050402}{DOI: 10.1103/PhysRevLett.87.050402.}

\bibitem{photon_wavefn0} B. Stoklasa, L. Motka, J. Rehacek, Z. Hradil and L. L. Sánchez-Soto, Wavefront sensing reveals optical coherence, Nature Commun. \textbf{5}, 3275 (2014).
\href{https://doi.org/10.1038/ncomms4275}{DOI: 10.1038/ncomms4275.}


\bibitem{photon_wavefn1} J. E. Sipe, Photon wave functions, Phys. Rev. A. \textbf{52}, 1875 (1995).
\href{https://doi.org/10.1103/PhysRevA.52.1875}{DOI: 10.1103/PhysRevA.52.1875.}

\bibitem{photon_wavefn2} I. B. Birula, Photon Wave Function, Progress in Optics \textbf{36}, 245-294 (1996).
\href{https://doi.org/10.1016/S0079-6638(08)70316-0}{DOI: 10.1016/S0079-6638(08)70316-0.}

\bibitem{tomo} B. J. Smith, B. Killett, M. G. Raymer, I. A. Walmsley, and K. Banaszek, Measurement of the transverse spatial quantum state of light at the single-photon level, Opt. Lett. \textbf{30}, 24, 3365-3367 (2005).
\href{https://doi.org/10.1364/OL.30.003365}{DOI: 10.1364/OL.30.003365.}


\bibitem{tomo2} A. I. Lvovsky and M. G. Raymer, 
Continuous-variable optical quantum-state tomography, Rev. Mod. Phys. \textbf{81}, 299 (2009).
\href{https://doi.org/10.1103/RevModPhys.81.299}{DOI: 10.1103/RevModPhys.81.299.}

\bibitem{tomo3} N. K. Langford, R. B. Dalton, M. D. Harvey, J. L. O’Brien, G. J. Pryde, A. Gilchrist, S. D. Bartlett, and A. G. White, Measuring Entangled Qutrits and Their Use for Quantum Bit Commitment, Phys. Rev. Lett. \textbf{93}, 053601 (2004).
\href{https://doi.org/10.1103/PhysRevLett.93.053601}{DOI: 10.1103/PhysRevLett.93.053601.}

\bibitem{tomo4} D. F. McAlister, M. Beck, L. Clarke, A. Mayer, and M. G. Raymer, Optical phase retrieval by phase-space tomography and fractional-order Fourier transforms, Opt. Lett. \textbf{20},1181-1183 (1995).
\href{https://doi.org/10.1364/OL.20.001181}{DOI: 10.1364/OL.20.001181.}


\bibitem{weak2} M. Mirhosseini, O. S. Magaña-Loaiza, S. M. H. Rafsanjani, and R. W. Boyd, Compressive Direct Measurement of the Quantum Wave Function, Phys. Rev. Lett. \textbf{113}, 090402 (2014).
\href{https://doi.org/10.1103/PhysRevLett.113.090402}{DOI: 10.1103/PhysRevLett.113.090402.}

\bibitem{transverseSpatialStructure} Z. Shi, M. Mirhosseini, J. Margiewicz, M. Malik, F. Rivera, Z. Zhu, and R. W. Boyd, Scan-free direct measurement of an extremely high-dimensional photonic state, Optica \textbf{2}, 388-392 (2015).
\href{https://doi.org/10.1364/OPTICA.2.000388}{DOI: 10.1364/OPTICA.2.000388.}

\bibitem{sagnac} C.C. Leary, L.A. Baumgardner, and M.G. Raymer, Stable mode sorting by two-dimensional parity of photonic transverse spatial states, Optics Express \textbf{17}, 2435-2452 (2009).
\href{https://doi.org/10.1364/OE.17.002435}{DOI: 10.1364/OE.17.002435.}



\bibitem{hongoumandel} C. K. Hong, Z. Y. Ou, and L. Mandel, Measurement of subpicosecond time intervals between two photons by interference, Phys. Rev. Lett. \textbf{59}, 2044-2046 (1987).
\href{https://doi.org/10.1103/PhysRevLett.59.2044}{DOI: 10.1103/PhysRevLett.59.2044.}


\bibitem{diff_microscopy} G. Popescu, T. Ikeda, Ramachandra R. Dasari, and M. S. Feld, Diffraction phase microscopy for quantifying cell structure and dynamics, Optics Letters \textbf{31}, Issue 6, 775-777 (2006).
\href{https://doi.org/10.1364/OL.31.000775}{DOI: 10.1364/OL.31.000775.}

\bibitem{microprin} D. Gabor, A new microscopic principle, Nature \textbf{161}, 777–778 (1948).
\href{https://doi.org/10.1038/161777a0}{DOI: 10.1038/161777a0.}

\bibitem{optholo} R. J. Collier, C. B. Burckhardt and L. H. Lin, Optical Holography (Academic, 1971).

\bibitem{OffaxisHolo} J. Mertz, Introduction to Optical Microscopy, 1st Edition, Roberts and company publishers, 217-220 (2019).

\bibitem{FiO} W. Szadowiak, S. Kundu, J. Szuniewicz, and R. Lapkiewicz, Self-referenced Measurement of the Spatial Structure of a Single Photon Beam, Frontiers in Optics + Laser Science APS/DLS, OSA Technical Digest (Optical Society of America, 2019), paper \textbf{FTu6A.6}.
\href{https://doi.org/10.1364/FIO.2019.FTu6A.6}{DOI: 10.1364/FIO.2019.FTu6A.6.}

\bibitem{Rochester} W. Szadowiak, S. Kundu, J. Szuniewicz, and R. Lapkiewicz, Self-referenced hologram of a single photon beam, Rochester Conference on Coherence and Quantum Optics (CQO-11), OSA Technical Digest (Optical Society of America, 2019), paper \textbf{W6A.27}.
\href{https://doi.org/10.1364/CQO.2019.W6A.27}{DOI: 10.1364/CQO.2019.W6A.27.}



\bibitem{hologramGen} N. Trautmann, G. P. Ferenczi, S. Croke, S. M. Barnett, Holographic quantum imaging: Reconstructing spatial properties via two-particle interference, Journal of Optics \textbf{19}, 5 (2017).
\href{https://doi.org/10.1088/2040-8986/aa67b6}{DOI: 10.1088/2040-8986/aa67b6.}


\bibitem{twophotonInterf1} L. Mandel, Quantum effects in one-photon and two-photon interference, Rev. Mod. Phys. \textbf{71}, S274 (1999).
\href{https://doi.org/10.1103/RevModPhys.71.S274}{DOI: 10.1103/RevModPhys.71.S274.}



\bibitem{singlePhotonInterference1} P. Grangier, G. Roger and A. Aspect, Experimental Evidence for a Photon Anticorrelation Effect on a Beam Splitter: A New Light on Single-Photon Interferences, Europhys. Lett. \textbf{1}, 173-179 (1986).
\href{https://doi.org/10.1209/0295-5075/1/4/004}{DOI: 10.1209/0295-5075/1/4/004.}



\bibitem{spdc2} C. E. Kuklewicz, M. Fiorentino, G. Messin, F. N. C Wong and J. H. Shapiro, High-flux source of polarization-entangled photons from a periodically poled ${\mathrm{KTiOPO}}_{4}$ parametric down-converter, Phys. Rev. A \textbf{69}, Issue 1, 013807 (2004).
\href{https://doi.org/10.1103/PhysRevA.69.013807}{DOI: 10.1103/PhysRevA.69.013807.}

\bibitem{IsCMOS} M. Jachura and R. Chrapkiewicz, Shot-by-shot imaging of Hong–Ou–Mandel interference with an intensified sCMOS camera, Opt. Lett. \textbf{40}, 1540–1543 (2015).
\href{https://doi.org/10.1364/OL.40.001540}{DOI: 10.1364/OL.40.001540.}

\bibitem{4f} J. Goodman, Introduction to Fourier Optics, 3rd Edition, Roberts \& Company Publishers, pages 248-249 (2005).

\bibitem{softwarePhotonfinder} R. Chrapkiewicz, W. Wasilewski, and K. Banaszek, High-fidelity spatially resolved multiphoton counting for quantum imaging applications, Opt. Lett. \textbf{39}, 5090-5093 (2014).
\href{https://doi.org/10.1364/OL.39.005090}{DOI: 10.1364/OL.39.005090.}

\bibitem{shearing1} W. J. Bates, A wavefront shearing interferometer, Proc. Phys. Soc. \textbf{59}, 940 (1947).
\href{https://doi.org/10.1088/0959-5309/59/6/303}{DOI: 10.1088/0959-5309/59/6/303.}

\bibitem{oam1} R. S. Aspden, M. J. Padgett, Light in a twist: Orbital angular momentum, Frontiers in Modern Optics \textbf{190}, 149 (2016).
\href{https://doi.org/10.3254/978-1-61499-647-7-149}{DOI: 10.3254/978-1-61499-647-7-149.}

\bibitem{oam2} S. M. Barnett, M. Babiker and M. J. Padgett, Optical orbital angular momentum, Phil. Trans. R. Soc. A \textbf{375}, 20150444 (2017).
\href{https://doi.org/10.1098/rsta.2015.0444}{DOI: 10.1098/rsta.2015.0444.}

\bibitem{SPIDER} A. Davis, V. Theil, M. Karpinski and B. J. Smith, Measuring the Single-Photon Temporal-Spectral Wave Function, Phys. Rev. Lett. \textbf{121}, 083602 (2018). 
\href{https://doi.org/10.1103/PhysRevLett.121.083602}{DOI: 10.1103/PhysRevLett.121.083602.}


\bibitem{matterWave} E. M. Rasel, M. K. Oberthaler, H. Batelaan, J. Schmiedmayer, and A. Zeilinger, Atom Wave Interferometry with Diffraction Gratings of Light, Phys. Rev. Lett. \textbf{75}, 2633 (1995).
\href{https://doi.org/10.1103/PhysRevLett.75.2633}{DOI: 10.1103/PhysRevLett.75.2633.}

\bibitem{matterWave2} M. Arndt, A. Ekers, W. Klitzing and H. Ulbricht, Focus on modern frontiers of matter wave optics and interferometry, New J. of Phys. \textbf{14}, 125006 (2012).
\href{https://doi.org/10.1088/1367-2630/14/12/125006}{DOI: 10.1088/1367-2630/14/12/125006.}



\end{thebibliography}

\begin{thebibliography}{99} 

\bibliographystyle{unsrt}
\bibitem{unwrap} M. A. Herráez, D. R. Burton, M. J. Lalor, and M. A. Gdeisat, Fast two-dimensional phase-unwrapping algorithm based on sorting by reliability following a noncontinuous path, Applied Optics \textbf{41}, Issue 35, pp. 7437-7444 (2002).
\end{thebibliography}
\end{document}